\documentclass[12pt]{iopart}

\usepackage{graphicx}  
\usepackage{wrapfig}   
\usepackage{subeqnar}  
\usepackage{threeparttable}

\begin{document}


\title[A global non-analytical fit method for the
complex lineshape analysis of Bi$_2$Sr$_2$CaCu$_2$O$_{8+\delta}$]
{A global non-analytical fit method for the complex lineshape
analysis of Bi$_2$Sr$_2$CaCu$_2$O$_{8+\delta}$}

\author{R. H. He\footnote[1]{Electronic mail: ruihuahe@stanford.edu; Present address:
Department of Applied Physics, Stanford University, Stanford,
California 94305} and  D. L. Feng}

\address{Physics Department, Applied Surface Physics
State Key Laboratory, and Synchrotron Radiation Research Center,
Fudan University, Shanghai 200433, China}


\begin{abstract}
The implementation of global non-analytical fit is exemplified by
its application to the spectral analysis of the complex lineshape
of Bi$_2$Sr$_2$CaCu$_2$O$_{8+\delta}$ at $(\pi,0)$. It deals with
properly a multi-level globality in the fitting parameters and
efficiently the non-analytical evaluation of the fitting function,
thus exhibiting a potential applicability to a wide range of
systematic analysis task.
\end{abstract}



Angle-resolved photoemission spectroscopy (ARPES) has greatly
advanced our microscopic understanding of the high-temperature
superconductivity by presenting high-precision measurements of
various dependence\cite{ZXReview}. Among them, for example, the
doping and temperature ($T$) dependent measurement on
Bi$_2$Sr$_2$CaCu$_2$O$_{8+\delta}$ (Bi2212) revealed the existence
of a coherent component in the lowest-lying binding energy
position which is closely related to the superconducting
transition\cite{FengScience}; the momentum and photo energy
($h\nu$) dependent investigations identified the long-sought
bilayer band splitting effect in the very overdoped (OD)
Bi2212\cite{FengPRL01} and argued for its persistence across a
large doping range of the sample\cite{FengPRB02}. Due to the
spectroscopic nature of ARPES, the intrinsic single-particle
spectral function is masked by the matrix element
effect\cite{MatrixElements} from a direct detection of the
multi-component lineshape, for example, the bilayer-split
\emph{peak-dip-hump} lineshape of the OD65 Bi2212 (with a $T_c$=65
K). In Ref.\cite{MyRapComm}, a combined analysis of the $h\nu$ and
$T$ dependence was achieved for a reliable extraction of the
intrinsic spectral information. To naturally account for the
physical constraints imposed by the intimate connection of the two
dependence of data, a global fitting program was developed, to our
knowledge, for the first time in the spectral analysis of complex
lineshape with simultaneous consideration on various dependence.

In this paper, we revisit this systematic spectral analysis on the
35 energy distribution curves (EDC's) [Fig. 1(a)-(c)] by detailing
the implementation of the global fitting procedure. It deals with
properly a multi-level globality (defined below) in the fitting
parameters and efficiently the non-analytical evaluation of the
fitting function, thus exhibiting a potential applicability to a
wide range of systematic analysis task.

\begin{figure}
\centerline{\includegraphics[width=3in]{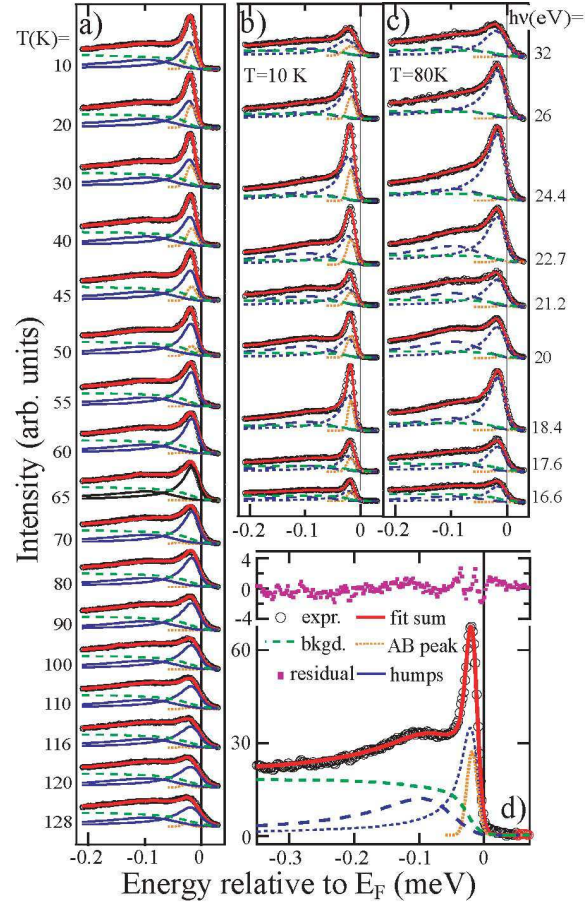}}
\caption[Fit results] {(color) Results of the global
non-analytical fit on the $T$-dependent (a) (see also [d] for an
enlargement of the $T$=30 K curves) and $h\nu$-dependent ARPES
spectra at $(\pi,0)$ of (Pb)-OD65 in the superconducting state at
$T$=10 K (b) and normal state at $T$=80 K (c). The AB hump lies
lower in energy than the BB hump. The width difference of the peak
is due to resolution variations from 10 to 18 meV at different
$h\nu$'s.} \label{fig1}
\end{figure}

For clarity, we recap the fitting function used for the lineshape
modelling in Ref. \cite{MyRapComm} with only a slight modification
described below.
\begin{eqnarray}
\hspace*{-2.5cm} \label{eq:Isum}
I(\omega,T,h\nu)&=&I_0(T,h\nu)\cdot[(\sum_{\alpha}^4J_{\alpha}(\omega,T,h\nu)\cdot
f(\omega,T))\otimes
R(\omega,\Gamma^\prime(h\nu))+B(\omega,T)]+I_1(T,h\nu), \nonumber
\end{eqnarray}
, where $f$ is the Fermi function, $R$ the $h\nu$-dependent
resolution Gaussian, $B$ the $T$-dependent empirical background
function\cite{MyRapComm}, $I_0$ and $I_1$ the EDC-specific linear
intensity coefficients. The summation is over the spectral
intensity of anti-bonding band (AB) hump, bonding band (BB) hump,
AB peak and BB peak, given by, respectively,
\begin{eqnarray}
\label{eq:I4}
J_{ah}(\omega,T,h\nu)&=&M_{ah}(h\nu)\cdot C_a(T)\cdot A_h(\omega,T,\alpha,\varepsilon_{ah}), \nonumber \\
J_{bh}(\omega,T,h\nu)&=&M_{bh}(h\nu)\cdot C_b(T)\cdot A_h(\omega,T,\alpha,\varepsilon_{bh}), \nonumber \\
J_{ap}(\omega,T,h\nu)&=&M_{ap}(h\nu)\cdot (1-C_a(T))\cdot A_p(\omega,\Gamma_a(T),\varepsilon_{ap}), \nonumber \\
J_{bp}(\omega,T,h\nu)&=&M_{bp}(h\nu)\cdot (1-C_b(T))\cdot
A_p(\omega,\Gamma_b(T),\varepsilon_{bp}) \nonumber
\end{eqnarray}
, where $M$ is the (squared) matrix element dependent only on
$h\nu$, $C$ the remaining spectral weight in the hump with its $T$
dependence to be fitted, $A_h$ and $A_p$ the self-normalized
spectral functions for the hump and the peak, respectively, given
by
\begin{eqnarray}
\label{eq:A}
A_h(\omega,T,\alpha,\omega_0)&=&\frac{\xi|\sum^{\prime\prime}(\omega,T)|}
    {(\omega-\sqrt{\omega_0^2+\Delta_{sc}(T)^2})^2+\sum^{\prime\prime}(\omega,T)^2},\nonumber \\
A_p(\omega,\Gamma(T),\omega_0)&=&\frac{2\sqrt{\ln2}}{\sqrt{\pi}\Gamma(T)}
    \exp[-(\frac{\omega-\omega_0}{\Gamma(T)/2})^2].\nonumber
\end{eqnarray}
We modified the empirical form of the imaginary part of
self-energy into
$\Sigma^{\prime\prime}(\omega,T)=|\sqrt{(\alpha\omega)^2+(\beta
T)^2}+\zeta|$, where lifetime broadening associated with the
impurity level of samples is reflected by $\zeta$ and that with
doping level by $\alpha$ and $\beta$. They are assumed to be
independent of $h\nu$ and $T$. $\xi$ is a self-normalization
factor given by
$\int_{-\infty}^{\infty}A_{h}(\omega)d\omega=1$\cite{ZXReview}.
$\Gamma(T)$ is a $T$-dependent linewidth of the peak. For the
bilayer-split bands in the superconducting state, a realistic SCG
opening introduces an effective $T$-dependent energy shift from
their $T$-independent renormalized band energy positions
($\varepsilon_{ah}$ and $\varepsilon_{bh}$) while the energy
position of the peak ($\varepsilon_{ap}$ or $\varepsilon_{bp}$) is
subjected to the fit.

In order to retrieve reliable quantitative information from a
multi-dimensional fit to a set of interconnected EDC's, many
physical constraints are involved and strictly followed. For
example, in the the $h\nu$-dependent set, all functions of $T$
($C$, for example) are shared variants to all EDC's at the same
$T$ while the functions of $h\nu$ (the matrix elements, for
example) are locally specified, and in the $T$-dependent set, vice
versa. In other words, for a given fitting parameter which is
subjected to a constrain, it is required to have the same value
for a specific group of EDC's (forming a so-called EDC's subset
for this parameter). Thus, the total number of its possible values
across the whole EDC set is smaller than the set size, 35. We
define this number the globality level ($Globality$ for this
fitting parameter. Particularly, the parameters with $Globality=1$
are in the top globality level, for example, $\varepsilon_{h}$,
$\alpha$ and those kept fixed during the fit, such as the
parameters for the BB peak component which is not considered in
the three-component fit exemplified below; the lowest globality
level ($Globality=35$) is owned by those local parameters with
their values specific to each EDC, such as $I_0$.

Technically, our fitting function contains some non-analytical
features, convolution by the resolution function and integration
in the self-normalization of spectral function. The conventional
point-by-point evaluation of the fitting function in the fitting
iteration fails to guarantee an efficient optimization to such a
large data set. We developed our fitting program in WaveMetrics
Igor Pro 4.0\cite{IgorNote1}. After implanting the fast algorithm
for non-analytical fitting\cite{IgorNote2} in the two-level global
fit procedure by Igor (where only the top and the lowest globality
levels are concerned), we further establish its multi-level global
fitting capability. We illustrate below the 5-step implementation
of our global non-analytical fit based on the 3-component model
(neglecting the $J_{bp}$ term in the first equation) on the $h\nu$
and $T$ dependent lineshapes of (Pb-)OD65\cite{MyRapComm}.

\begin{enumerate}
\item[Step 1]: Include in the fit the 9 normal state ($@80K$) then
the 9 superconducting state ($@10K$) $h\nu$-dependent EDC's on
Pb-OD65 (in an ascending order sorted by the $h\nu$ values,
respectively). Append to the EDC's set the 17 $T$-dependent EDC's
on OD65 using He-I (in a descending order sorted by the $T$
values). The EDC's are labelled from 0 to 34 and concatenated
sequentially in X and Y scale, respectively, to form a huge "EDC".

\begin{table*}[t]
\parindent=0pt
\hspace*{-0.5cm}
\begin{threeparttable}
\vspace*{-0.3cm}\caption{List of fitting parameters}{\small
\begin{tabular}[b]{l c||ccc ccc ccc ccccc ccc}
\hline \hline
 Para No. &&& 0 & 1 & 2 & 3 & 4\\
\hline
 Parameters &&& $I_0$ & $M_{ah}$ & $M_{bh}$ & $M_{ap}$ & $M_{bp}$ \\
\hline
 $Globality$\tnote{b} &&& $35$ & $10(2\times 9+17\times 1)$ & $10(2\times 9+17\times 1)$
 & $10(2\times 9+17\times 1)$ & $1$ \\
\hline
 Fixation &&& $\times$ & $\times$ & $\times$ & $\times$ & arb.\tnote{a} \\
\hline
\end{tabular}
\begin{tabular}[b]{l c||ccc ccc ccc ccccc ccc}
\hline \hline
 Para No. && 5 & 6 && 7 && 8 & 9 & 10 & 11  \\
\hline
 Parameters && $C_a$ & $C_b$ && $\varepsilon_{ah}$ && $\varepsilon_{bh}$ & $\varepsilon_{ap}$ & $\varepsilon_{bp}$ & $\alpha$  \\
\hline
 $Globality$ && $17(10\times 2+1\times 15)$ & $1$ && $1$ && $1$ & $17(10\times 2+1\times 15)$ & $1$ & $1$ \\
\hline
 Fixation && $\times$ & 1 && $\times$ && $\times$ & $\times$ & arb. & $\times$ \\
\hline
\end{tabular}
\begin{tabular}[b]{l c||ccc ccc ccccc ccccc ccccc ccc ccc}
\hline
 Para No. && 12 && 13 && 14 && 15 && 16 && 17\\
\hline
 Parameters && $\Gamma_a$ && $\Gamma_b$ && $\delta E_F$\tnote{d}
  && $I_1$ && $\beta$ && $\zeta$\\
\hline
 $Globality$ && $2(18\times 1+17\times 1)$\tnote{c} && $1$ && $1$ && $35$ && $1$ && $2(18\times 1+17\times 1)$\tnote{c}\\
\hline
 Fixation && $\times$ && arb. && $0$ && $\times$ && $\times$ && $\times$\\
\hline \hline
\end{tabular}}
\begin{tablenotes}
 \scriptsize
\item[a] arbitrary value; \item[b] the number without (or outside)
the bracket: $Globality$ (number of subsets); expression inside
the bracket: size of each subset (number of EDC's contained)
$\times$ number of subsets of the same size; \item[c] to assume
that the quality difference between Pb-OD65 and OD65 used in the
two dependent investigations should be reflected \emph{only} in
the quasi-particle lifetime; \item[d] a slight correction in
energy for $E_F$ of each EDC to account for the uncertainty in
$E_F$ alignment.
\end{tablenotes}
\end{threeparttable}
\label{Tab1}
\end{table*}

\item[Step 2]: Discriminate the fitting parameters of the top
globality level from the otherwise (of intermediate globality or
the lowest globality). Then fix those parameters if they are
required to be invariable during the fit. We summarize in Table 1
the properties of all the fitting parameters involved.

\item[Step 3]: Build a global fit parameters set, which includes
sequentially, the 9 parameters with $Globality=1$, the remaining 9
parameters with $Globality>1$ ($9\times 35=315$, totally) grouped
in the number order of the EDC in the EDC concatenation to which
they belong. Hence, we end up with another index to locate the
parameter in a global manner, called GL Para No. hereafter, from
which we can set up a bidirectional mapping to the original Para
No. (in Table 1) for a specific EDC.

\begin{figure}[t]
\centerline{\includegraphics[width=5in]{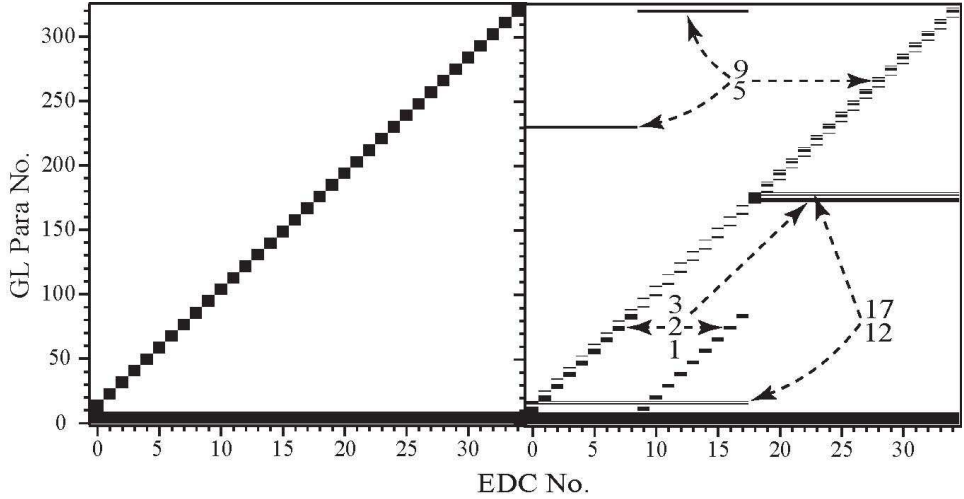}}
\caption[Truth Table] {Truth tables as black-and-white plots
(black: True, white: False), where the true pertinence of elements
in the global fit parameters set to each EDC is recovered for a) a
two-level global fit or b) a multi-level global fit in the example
problem. The arrows identified by the Para No. locate the
corresponding features of the fitting parameters with intermediate
globality levels. For example, the three arrows from "1" indicate
$M_{ah}$ has two independent values for the $h\nu$ and $T$
dependence subset, respectively, one for the EDC pairs taken at
the same $h\nu$ and the other for all the EDC's taken with He-I
(the so-called parameter degeneracy of $M_{ah}$). A horizontal
line segment with unit length stands for a parameter with
$Globality=35$ while a full-scale line segment for the one with
$Globality=1$. Note that the patterns may vary dramatically with
different orders the EDC and parameters of the sets would be
arranged in.} \label{fig2}
\end{figure}

\item[Step 4]: The global fit problem on a set of EDC's has been
converted into a fit on the EDC concatenation with the global fit
parameters set. Having in mind the unique calling pattern of the
fitting function in Igor curve fitting routine\cite{IgorNote2} and
the time-consuming non-analytical evaluation in the fitting
function (see Ref. \cite{MyRapComm}), a direct fit is practically
unfeasible, which asks for a function calling time up to
$(M+1)\times N$ ($M$ and $N$ denote the total GL Para No. and
point numbers of the EDC concatenation, here $324$ and $\sim5400$,
respectively) for a single iteration! The introduction of a fast
algorithm greatly reduces the function calling time down to
$(M+1)\times L$ ($L$ denotes the total EDC number included in the
concatenation, here 35) at the expense of a size increase in the
RAM which is used for the storage of the trial fit
curves\cite{IgorNote2}. We can further cut down the time to
$(K+1)\times L$ ($K$ denotes the total Para No., here 18) by
exploiting the fact that in a global fit a large number of the
global fit parameters are actually \emph{not} related to the
function evaluation for a given EDC, which needs to be reduced
from the EDC concatenation during each iteration of the fit. This
is because only $K$ out of $M$ parameters in the global fit
parameters set comes from a single EDC.

To pick out properly those relevant parameters which really
contribute to the function update on a given EDC, a truth table
helps, which is convenient to build at the end of Step 3. In Fig.
\ref{fig2}(a), we show this binary table as a black-and-white
plot. The bottom black wall and black bricks in the diagonal
clearly indicate the order in which the global fit parameters set
is organized and, especially, the EDC-specific grouping nature of
the $Globality>1$ parameters following the $Globality=1$ ones
(reminiscent of Step 3).

\item[Step 5]: The global fit involves only two levels in Step 4.
The task of this step is to achieve a multi-level globality
defined in Table 1. This \emph{only} requires our making
modifications on the truth table accordingly. Fig. \ref{fig2}(b)
shows a truth table modified to achieve the wanted globality
configuration. In contrast to Fig. \ref{fig2}(a), the introduction
of intermediate globality levels to the global fit parameters set
results in the parameters degeneracies for certain EDC's as well
as, in return, some "void" parameters, which seemingly have no
effect on \emph{any} EDC in the concatenation. To prevent Igor
curve fit routine from "a singular matrix error" report, we simply
fix them to arbitrary values during the fit. This modification
doesn't alter the conservation in the column sum of truth values
in the table (i.e., the total Para No., $K$), and thus neither the
iteration efficiency.
\end{enumerate}

In Fig. \ref{fig1}, the fit curves collapse well onto the
experimental EDC's, reminiscent of Figs. 2-3 in
Ref.\cite{MyRapComm}. Pay attention to the great contrast in the
overall quality (or details) of the fit between the ones shown in
Fig. 2 of Ref. \cite{Fink2} and ours, though on data sets
generally twice larger in size.  This suggests the local fit
scheme\cite{Fink2,Fink4} with manual adjustments on the global
fitting parameters is unlikely to yield a reliable result in a
truly global sense. In contrast, our fully-automatic fitting
routine can yield a robust\cite{RobustFit}, efficient and
physically-constrained global fit for a large data set, which is
frequently encountered in the spectral analysis of ARPES.

Note that the algorithm proposed here is \emph{not} limited to the
application of the ARPES lineshape analysis only, but of a great
applicability to a wide range of problems where a multi-level
global analysis or/and non-analytical nonlinear fit is involved.
For example, the analysis of hysteresis curves in the magnetism
research can be a good case in point for its potential
application. The angle-dependent hysteresis measurement contains
the information of magnetic anisotropy which is currently
retrieved only \emph{qualitatively} due to the lack of a global
fit on the angle-dependent curves set. Furthermore, in most cases,
the single-domain coherent rotation model in the textbook is not
sufficient to include the necessary physics to reproduce the
hysteresis curve in calculation. By using spin
dynamics\cite{SpinDyn} as the non-analytical evaluation kernel in
the global fit, the roles of collaborative spin motion and domain
formation which are requisite for the hysteresis can be
investigated quantitatively. Generally, our algorithm provides a
useful and universal guideline to serve the advanced systematic
analysis in experimental physics.

D.L.F. is supported by the NSFC grants 10225418 and 10321003, and
Shanghai municipal Phosphor Project. SSRL is operated by the DOE
Office of Basic Energy Science Division of Material Sciences.



\begin{thebibliography}{}
\bibitem{IgorNote1} Igor uses the Levenberg-Marquardt algorithm to search for the
coefficient values that minimize $\chi^2$. This is a form of
non-linear, least-squares fitting. This kernel is well-packed and
convenient for customizing interfaces oriented for different uses.
\end{thebibliography}
\end{document}